\documentclass[prb,aps,amssymb,showpacs,twocolumn]{revtex4}
\usepackage{amsmath}
\usepackage{amssymb}
\usepackage{amsthm}
\usepackage{amsfonts}
\usepackage{enumerate}
\usepackage{latexsym}
\usepackage{graphicx}
\usepackage{dcolumn}
\usepackage{bm}
\usepackage{natbib}
\usepackage{citesort}

\newcommand{\beq}{\begin{equation}}
\newcommand{\eneq}{\end{equation}}

\input{epsf}

\begin{document}

\tolerance 10000

\newcommand{\vk}{{\bf k}}

\preprint{APS/123-QED}

\title{Non-diffusive spin dynamics in a two-dimensional electron gas}
\author{C.P. Weber$^1$}
\email{cpweber@lbl.gov}
\author{J. Orenstein$^1$}
\author{B. Andrei Bernevig$^2$}
\author{Shou-Cheng Zhang$^2$}
\author{Jason Stephens$^3$}
\author{D.D. Awschalom$^3$}
\affiliation{$^1$Physics Department, University of California,
Berkeley and \\ Materials Science Division, Lawrence Berkeley
National Laboratory, Berkeley, CA 94720 \\ $^2$Physics Department,
Stanford University, Stanford, CA 94305 \\ $^3$Center for Spintronics and Quantum Computation,
University of California, Santa Barbara, California 93106, USA}%
\date{\today}

\begin{abstract}
We describe measurements of spin dynamics in the two-dimensional
electron gas in GaAs/GaAlAs quantum wells.  Optical techniques,
including transient spin-grating spectroscopy, are used to probe
the relaxation rates of spin polarization waves in the wavevector
range from zero to $6\times 10^4$ cm$^{-1}$.  We find that the
spin polarization lifetime is maximal at nonzero wavevector, in
contrast with expectation based on ordinary spin diffusion, but in
quantitative agreement with recent theories that treat diffusion
in the presence of spin-orbit coupling.
\end{abstract}

\date{\today}


\maketitle

Electronic systems with strong spin-orbit interaction (SOI)
exhibit exotic effects that arise from the coupling of spin
polarization and charge current, such as spin Hall currents
\cite{d'yakonov1971,hirsch1999,murakami2003,sinova2004,KatoSpinHall,WunderlichSpinHall,sih2005}
and current-induced spin polarization \cite{kato2004a}. These
effects involve manipulation of the electron spin via electric,
rather than magnetic, fields, creating the potential for
applications in areas from spintronics to quantum computing
\cite{wolf2001}. However, SOI is a double-edged sword, as it also
has the undesired effect of causing decay of spin polarization,
reflecting the nonconservation of the total spin operator,
$\vec{S}$, i.e. $[\vec{S},\mathcal{H}] \ne 0$, where $\mathcal{H}$
is any Hamiltonian that contains spin-orbit coupling.

In the consideration of spin-orbit coupled systems, a great deal
of attention has been focused on quantum wells or heterostructures
fabricated in III-V semiconductors, where breaking of inversion
symmetry allows coupling that is linear in momentum.  For
electrons propagating in [001] planes, the most general form of
the linear coupling includes both Rashba \cite{bychkov1984} and
Dresselhaus \cite{dresselhaus1955} contributions:

\begin{equation}
\mathcal{H}_{so}=\alpha (k_y \sigma_x-k_x\sigma_y)+\beta
(k_x\sigma_x-k_y\sigma_y), \label{hamiltonian}
\end{equation}

\noindent where $k_{x,y}$ is the electron wavevector along the
[10], [01] directions in the plane, and $\alpha$ and $\beta$ are
the strengths of the Rashba and Dresselhaus couplings.  The
spin-orbit terms generate an effective in-plane magnetic field,
$\vec{b}_{so}$, whose direction depends on $\vec{k}$. Spin
nonconservation takes the form of precession, at a rate governed
by $\vec{b}_{so}(\vec{k})$, during an electron's free flight
between collisions.  In the Dyakanov-Perel (DP) regime
\cite{Dyakonov1972}, where the precession angle during the
free-flight time $\tau$ is small, a spatially uniform spin
polarization will relax exponentially to zero at the rate
$1/\tau_s \propto |\vec{b}_{so}|^2 \tau$. Although this process
relaxes the initial polarization state, spin memory is not
entirely lost, even after the electron undergoes many collisions.
The relationship between real-space trajectory and spin precession
leads to a correlation between the electron's position and its
spin. Such correlations are predicted to enhance the lifetime of
certain spatially inhomogeneous spin polarization states beyond
what would be expected for conventional spin diffusion
\cite{BurkovMacDonald}.

Burkov \textit{et al.} \cite{BurkovMacDonald} considered
$\mathcal{H}_{so}$ with only Rashba coupling ($\beta=0$), and
showed that a helical spin density wave with wavevector
$\sqrt{15}m\alpha/2$ decays more slowly than a uniform (or $q=0$)
spin polarization, by a factor $32/7$. This contrasts with
ordinary spin diffusion, where the decay rate of a polarization
wave increases monotonically ($\propto q^2$) with increasing
wavevector. More recently, Bernevig \textit{et al.}
\cite{bernevig2006} showed that the lifetime of the spin helix is
enhanced further in the presence of both Rashba and Dresselhaus
coupling, and diverges as $\alpha\rightarrow\beta$. The infinite
lifetime, or "persistent spin helix" (PSH) state, is a
manifestation of an exact $SU(2)$ symmetry of $\mathcal{H}_{so}$
at the point $\alpha=\beta $. As a consequence of the $SU(2)$
symmetry, a spiral spin polarization in the $z,x_+$ plane (where
$\hat{z}$ and $\hat{x}_+$ are the normal and [11] directions,
respectively) with wavevector $4m\alpha \hat{x}_+$ is a conserved
quantity. These predictions suggest the possibility of
manipulating spin polarization through SOI, without necessarily
compromising spin memory, by controlling $\alpha/\beta$ with
externally applied electric fields.

Transient spin grating (TSG) experiments \cite{Cameron,Weber2005}
are particularly well-suited to testing the theoretical
predictions of Refs. \cite{BurkovMacDonald,bernevig2006}, as they
directly probe the decay rate of nonuniform spin distributions. In
the TSG technique, a spatially periodic polarization of the
out-of-plane spin, $S^z$, is created by a pair of interfering
ultrashort laser pulses.  By varying the relative angle of the two
pump beams, we are able to vary the magnitude of the grating
wavevector $\vec{q}$ in the range $0.44-5.3\times10^4$ cm$^{-1}$,
corresponding to wavelengths in the range from $\approx 1-15$
microns. The subsequent time evolution of the spin polarization
wave amplitude, $S^z_q(t)$, is measured by coherent detection of a
time-delayed probe beam that diffracts from the photoinjected
"spin grating."

In this work we report detailed TSG measurements on two samples,
each consisting of 30 quantum wells of thickness 120 $\AA$. One
was with $\delta$-doped with Si in the barriers to create a high
mobility ($\mu\simeq 150,000$ cm$^2$/V-s at 5 K) two-dimensional
electron gas (2DEG) with carrier density 7.8 $\times 10^{11}$
cm$^{-2}$. The $\delta$ layers are symmetrically located on either
side of each well. The second sample is identical except that the
mobility of the electrons was reduced to $\simeq 3,500$ cm$^2$/V-s
by placing 83$\%$ of the dopant atoms in the wells, rather than in
the barriers.

The TSG measurements show that the lifetime of $S^z_q$ is maximal
at nonzero $|\vec{q}|$ in both samples, in contrast to
expectations for simple diffusion but in agreement with the
prediction of Burkov \textit{et al.} \cite{BurkovMacDonald}. The
magnitude of the lifetime enhancement and its dependence on the
direction of $\vec{q}$ point to the presence of both Rashba and
Dresselhaus interactions in our samples. Quantitative agreement
between the theory of Bernevig et al. \cite{bernevig2006} and our
experimental results suggests the feasibility of realizing and
detecting a PSH in samples engineered to achieve
$\alpha\approx\beta$.

\begin{figure}[h]
\includegraphics[scale=.4]{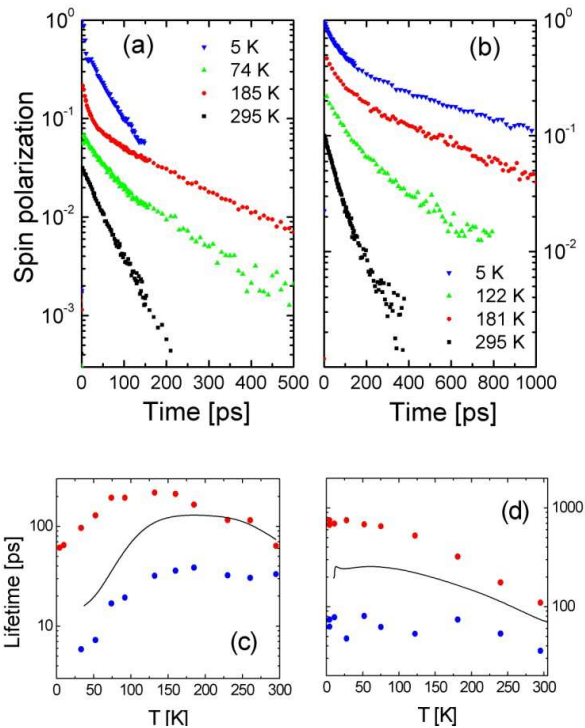}
\caption{[Color Online] Decay of the out-of-plane component of
spin polarization wave with wavevector in the [11] direction, for
(a) the high-mobility sample with $|\vec{q}|=0.58\times10^4$
cm$^{-1}$ and (b) the low-mobility sample with
$|\vec{q}|=0.69\times10^4$ cm$^{-1}$, for several temperatures.
The initial value of spin polarization is estimated at a few
percent for all measurements. Decay curves are offset vertically
for clarity. Solid symbols in panels (c) and (d) indicate the
temperature dependence of the two time constants obtained from a
double exponential fit to the decay curves. Solid lines are the
lifetime of a spatially uniform spin polarization.}
\label{timetraces_2_samples}
\end{figure}

Panels (a) and (b) of Fig. 1 show the decay of $S_q^z(t)$ for the
high-$\mu$ and low-$\mu$ samples, respectively, measured at the
wavevectors where the decay rate is smallest. In both samples the
decay of the grating is clearly not a single exponential at low
$T$ and crosses over to nearly single exponential as room
temperature is approached. All the decay curves can be fitted to a
double exponential form, $a_1\exp(-t/\tau_1)+a_2\exp(-t/\tau_2)$,
with equal weighting factors ($a_1=a_2$) over almost the entire
range of $T$. The only exception is the $T<25$ K regime of the
high-$\mu$ sample, where the momentum relaxation rate, $1/\tau$,
becomes comparable to the spin precession frequency, $\Omega$.  In
this ($\Omega\tau\geq 1$) regime the initial decay is damped
oscillatory rather than exponential \cite{Weber2005}. The solid
symbols in panels (c) and (d) are the time constants $\tau_1$ and
$\tau_2$ extracted from the double-exponential fit, plotted as a
function of $T$. For both samples the ratio between the fast and
slow rates is at least a factor of ten at low $T$ and gradually
diminishes as $T$ approaches room temperature. The solid lines are
the lifetimes of the uniform spin polarization, $\tau_s(T)$, as
obtained from decay of transient Faraday rotation induced by a
single, circularly polarized pump beam. For all temperatures, the
$q=0$ polarization decays as a \textit{single} exponential, whose
lifetime lies \textit{between} the two lifetimes observed at
nonzero $q$.

The dispersion of the decay rates with $\vec{q}\|[11]$ is shown in
Figs. 2(a) and 2(b), for the high and low $\mu$ samples,
respectively. We present results for $T=50$ K, a temperature at
which both samples are in the $\Omega\tau<1$ regime, and yet the
ratio $\tau_2/\tau_1$ remains large (in a certain range of $q$).
For both samples, the larger of the two time constants peaks
sharply at $|\vec{q}|\approx 0.6\times 10^4$ cm$^{-1}$. The
lifetime of the more rapidly decaying component decreases
monotonically with increasing $|\vec{q}|$, consistent with simple
diffusion.

\begin{figure}[h]
\includegraphics[scale=0.4]{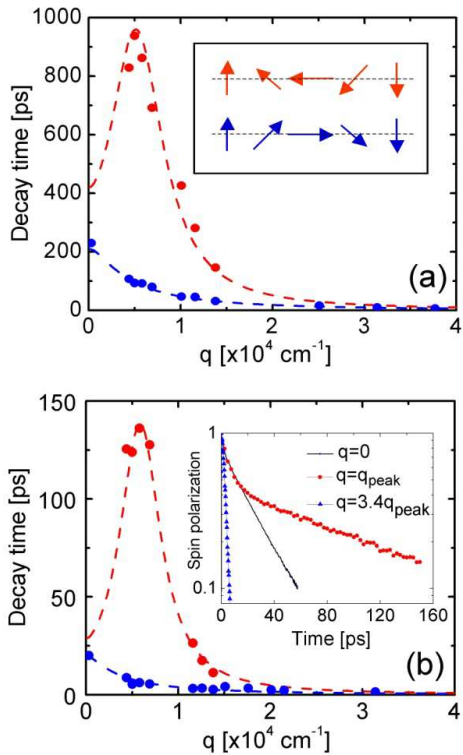}
\caption{ [Color Online] Decay lifetimes obtained from best fit of
a double exponential to experimentally measured decay of spin
polarization, as a function of wavevector $\vec{q}$ parallel to
the $[11]$ crystal axis. Data are shown for the (\textbf{a})
low-$\mu$ and (\textbf{b}) high-$\mu$ samples, at 50 K.  The
dashed lines are fits to a theory (described in the text) in which
the two lifetimes correspond to the positive and negative spin
helices illustrated in\textbf{ Fig. 2(a) inset}.  \textbf{Fig.
2(b) inset:} Comparison of spin polarization decay at $q=0$,
$0.58\times 10^4$ cm$^{-1}$, and $2.01\times 10^4$
cm$^{-1}$.}\label{dispersion_curves}
\end{figure}

The spin-dynamical effects illustrated in Figs. 1 and 2,
biexponential decay and non-monotonic dispersion, are in
quantitative agreement with the theory of coupled charge and spin
dynamics in the presence of $\mathcal{H}_{so}$, which we describe
briefly below. Assuming a single, isotropic $\tau$, Burkov
\textit{et al.} \cite{BurkovMacDonald} and Bernevig \textit{et
al.} \cite{bernevig2006} derived a set of four equations that
describes the coupling of electron density $n_q(t)$ and the three
components of $\vec{S}_q(t)$ brought about by the SOI. Along the
$[11]$ and $[1\bar{1}]$ directions, the four equations separate
into two coupled pairs. For $\vec{q}\parallel[11]$, spin precesses
in the $z-x_+$ plane, leading to coupling of $S^{x+}$ and $S^z$.

Solving the pair of equations that couple $S^z$ and $S^{x+}$
yields two eigenmodes and corresponding eigenfrequencies
$i\omega_{1,2}(q)\equiv 1/\tau_{1,2}(q)$. The $\vec{q}\parallel
[11]$ eigenfrequencies are,

\begin{eqnarray}
\frac{1}{\gamma_0\tau(q)} & = & \frac{q^2}{2q_0^2}+3+\sin2\phi
\nonumber\\ & \pm &
\sqrt{(1-\sin2\phi)^2+\frac{4q^2}{q_0^2}(1+\sin2\phi)},
\label{eq:eigenfrequencies}
\end{eqnarray}

\noindent where $q_0 \equiv m^\ast v_{so}/\hbar$ is the reciprocal
of the distance required for an electron to precess by $2\pi$ in
the spin-orbit field and $\gamma_0\equiv v_{so}^2 k_F^2 \tau$ is
the typical DP (\textit{\textit{i.e.}}, $q=0$) decay rate. The
parameters $\phi \equiv\tan^{-1}(\alpha/\beta)$ and $v_{so}\equiv
\sqrt{\alpha^2+\beta^2}$ reflect the relative and combined
coupling strengths of the Rashba and Dresselhaus interactions,
respectively. Finally, the spin diffusion coefficient $D_s\equiv
v_F^2\tau/2$ is given by $\gamma_0/2q_0^2$.

The solutions for $\vec{q}\|[11]$ are especially interesting at
the $SU(2)$ point where $\alpha=\beta$, and $\sin2\phi=1$. In this
case the eigenvectors are $(1, \pm i)/\sqrt{2}$, corresponding to
forward and backward spin spirals in the $z,x_+$ plane, for each
value of $q$. A TSG experiment creates the initial condition
$(1,0)$, which couples with equal strength to the two
eigenvectors. The subsequent time evolution of the two eigenmodes
has the form of a double exponential decay with equal weighting
factors. At the ``resonant'' wavevector $q_0\sqrt{8}$, the two
decay rates are 0 and $16\gamma_0$. After decay of the unstable
eigenmode, the PSH remains stable, despite the rapid electron
scattering and spin precession that are occurring. The
eigenfrequenies for $\vec{q}\|[1\bar{1}]$ can be obtained from Eq.
\ref{eq:eigenfrequencies} as well, if we replace $\phi$ by
$-\phi$. At the same $SU(2)$ point where the PSH exists for
$\vec{q}\|[11]$, the spin dynamics for $\vec{q}\|[1\bar{1}]$ obey
simple diffusion.

Several features of the physics at the $SU(2)$ point remain when
$\alpha\neq\beta$, but both are nonzero.  The lifetime at the
resonant wavevector is enhanced relative to the case when only one
of two interactions is present, although it no longer diverges.
The enhancement is expected to be stronger for $\vec{q}\|[11]$
than $\vec{q}\|[1\bar{1}]$. The eigenvectors are still admixtures
of $S^z$ and $S^{x+}$ and the photoinjected wave of pure $S^z$
will again decay as a double exponential. However, the weighting
factors are $q$-dependent and the long-lived state will be an
elliptical, rather than circular, spin helix.

With this overview of the theory, we return to the experimental
data.  The dotted lines in Fig. 2 show the best fits obtained by
varying parameters in Eq. \ref{eq:eigenfrequencies}. The fits
yield parameter values $v_{so}=460$ m/s, $\phi=0.2$, and
$D_s=1000$ cm$^2$/s for the higher-$\mu$ sample and $v_{so}=480$
m/s, $\phi=0.08$, and $D_s=105$ cm$^2$/s for the intentionally
disordered sample. The above value of $v_{so}$ for the high-$\mu$
sample predicts a spin-precession frequency for ballistic
electrons, $\Omega=v_{so}k_F=0.10 $ THz, that is within 10$\%$ of
the experimental value \cite{weberthesis}. The theory applies
equally well to both samples, despite a difference of about ten in
their scattering rates. Note that smaller $\tau$ leads to smaller
spin relaxation rates, as the entire dispersion scales with the DP
relaxation rate, $4\gamma_0$.

A surprising feature of the theory \cite{bernevig2006} is the
sensitivity of the dispersion curves to small admixtures of Rashba
coupling into a pure Dresselhaus system, or \textit{vice versa}.
Our quantum wells are designed to be perfectly symmetric and
therefore no Rashba term is expected. However, the maximum
lifetime enhancement, $\sim 7$, for the higher $\mu$ sample is
significantly larger than the factor 32/7 predicted for
$\alpha=0$. This discrepancy is accounted for by a relatively
small admixture $\alpha\simeq 0.2\beta$. It is possible that the
presence of a small Rashba contribution can be traced differences
in the interface on either side of the well, a known feature of
the GaAs/GaAlAs system \cite{interfaceroughness,deandrada1997}. It
is interesting that the value $\alpha/\beta\simeq0.08$ is smaller
in the lower-$\mu$ sample. The difference may reflect the tendency
of the ionized impurities to attract electrons toward the barrier
in case of $\delta$-doping, as opposed to doping inside the well.

\begin{figure}[h]
\includegraphics[scale=.4]{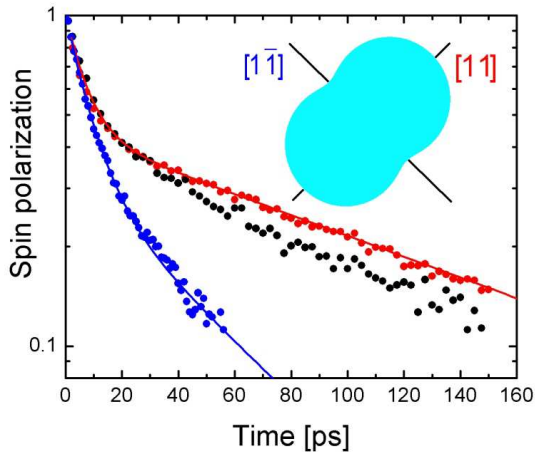}
\caption{ [Color Online] Decay of spin polarization in the
high-$\mu$ sample as a function of time for three orientations of
the grating wavevector (top-to-bottom: $[11]$, $[10]$, and
$[1\bar{1}])$. The lines through the data are predictions of the
theory described in the text, using the parameters that were
obtained from fitting the dispersion curves. The anisotropy is
smaller in the low-$\mu$ sample, which is consistent with the
conclusion based on the dispersion data. \textbf{Inset:} Fermi
contour for a system of free electrons subject to the SOI
described by Eq. \ref{hamiltonian}, showing the origin of the
extremal orientations. Rashba coupling set to one-half of
Dresselhaus (larger than in the systems under study here) for the
purpose of illustration.} \label{anisotropy}
\end{figure}

A further prediction of the theory is that even small
$\alpha/\beta$ generates a large anisotropy in the dispersion
curves. To test this prediction we measured $S_q^z$ at 50 K in the
higher-$\mu$ sample, for $\vec{q}$ oriented along the $[11]$,
$[1\bar{1}]$, and $[10]$ directions. The results shown in Fig.
\ref{anisotropy} demonstrate that the lifetime enhancement is
strong for $\vec{q}\|[11]$ and weak for $\vec{q}\|[1\bar{1}]$. The
line through the $[11]$ curve is calculated using Eq.
\ref{eq:eigenfrequencies} for the lifetimes and Eq. 22 of Ref.
\cite{bernevig2006} for the weighting factors of the double
exponential decay. The parameters are the same as used to fit the
dispersion along $[11]$. The weighting factors and decay rates for
$\vec{q}\parallel[1\bar{1}]$ are calculated using the \textit{same
parameters}, provided that we replace $\phi$ by $-\phi$. The
theory is clearly quite successful in providing a quantitative
description of the spin relaxation dynamics seen in our
experiment.

The only feature of the data not captured by the theory described
above is the gradual decrease in the $\tau_2/\tau_1$ ratio with
increasing $T$ (shown in Fig. \ref{timetraces_2_samples}) that
takes place for both samples. The characteristic scale of $T$
clearly is not the spin-orbit splitting, which is $\sim v_F q_0$
or about 1 K for our samples. We speculate that the $T$-dependence
of $\tau_2/\tau_1$ is a consequence of the cubic (in $k$)
Dresselhaus coupling, $\mathcal{H}_{cD} \propto k_x k_y^2\sigma_x
- k_y k_x^2 \sigma_y$, which does not appear in Eq.
\ref{hamiltonian}. $\mathcal{H}_{cD}$ is non-negligible when
$\langle k_z^2 \rangle$, the expectation value of $k_z^2$ in the
lowest subband, becomes comparable to $k_F^2$. The cubic
Dresselhaus coupling destroys the proportionality $\Omega\propto
k$, which ensures that the precession angle in a free-flight
between collisions depends only on the electron's displacement,
and not on its velocity. In the presence of $\mathcal{H}_{cD}$ the
precession angle will depend on $k$, and the fractional spread in
precession angle for a given $\Delta k$ is $\sim k_F \Delta k
/\langle k_z^2\rangle$. Assuming a thermally induced momentum
distribution $\Delta k \sim T/v_F$, the relative variation in
precession angle is $\sim T/E_1$, the ratio of the temperature to
the energy of the first excited subband. Our results are
consistent with $E_1$ (which is $\approx 500$ K for our samples)
as the characteristic energy scale above which the correlations
that generate the PSH are lost.

Finally, we discuss directions for future research that are
suggested by the conclusions of this work. The main implication of
the theory and experiments described above is that $SU(2)$ spin
symmetry can be recovered, despite the presence of SOI, if the
condition $\alpha=\beta$ can be satisfied. This can be
accomplished by increasing the strength of the Rashba interaction,
as compared with our samples, by designing asymmetric doping
and/or barrier profiles, or by applying a gate electrode. Once
achieved, the $SU(2)$ symmetry is robust with respect to disorder
and interactions \cite{bernevig2006}, and could be stabilized at
room temperature by increasing $E_1$. Schliemann \textit{et al.}
\cite{egues2003} previously discussed some attractive features of
a system in which $\alpha=\beta$ for logic devices based on spin
transport. They described a device structure in which spin is
injected and detected at two points, and showed that if
$\alpha=\beta$ the change in direction of the injected spin
depends only on the vector displacement, $\vec{a}$, of the two
point contacts, even in the presence of strong scattering.
Modulating the Rashba coupling via an external gate can then
destroy the long-range spin correlation. The restriction to point
contacts is actually unnecessary in the $SU(2)$ system because the
net spin precession depends only on the projection of $\vec{a}$
onto $\hat{x}_+$. For example, spin injected with arbitrary
polarization from anywhere along a line contact defined by $x_+
=0$ will be detected with the same spin anywhere along a line
contact at $x_+=n\pi/2m\alpha$, where $n$ is an integer. Thus, in
principle, a planar spin logic device operating at room
temperature is feasible, if highly spin selective contacts to the
2DEG can be fabricated.

B.A.B. wishes to acknowledge the hospitality of the Kavli
Institute for Theoretical Physics at University of California at
Santa Barbara, where part of this work was performed. This work is
supported by the NSF through the grants DMR-0342832, DMR-0305223,
and by the US Department of Energy, Office of Basic Energy
Sciences under contract DE-AC03-76SF00515.

\end{document}